\documentclass[a4paper,twocolumn,prb,superscriptaddress,showpacs]{revtex4}
\usepackage{dcolumn}
\usepackage{amsmath}
\usepackage{graphicx}
\usepackage{latexsym}  
\usepackage{amsfonts}  
\usepackage{amssymb}  
\DeclareGraphicsExtensions{.eps,.pdf,.png,.gif,.jpg}  
  
\newcommand{\be}{\begin{equation}}
\newcommand{\ee}{\end{equation}}
\newcommand{\bea}{\begin{eqnarray}}
\newcommand{\eea}{\end{eqnarray}}
\newcommand{\Id}[1] {\int \! \! {\rm d}^3 #1}
\newcommand{\ID}[1] {\int \! \! \! \frac{{\rm d}^3 #1}{(2 \pi)^3}}
\renewcommand{\vr} {{\bf r}}
\newcommand{\ve} {{\mathbf e}}
\newcommand{\vm} {{\mathbf m}}
\newcommand{\vk} {{\mathbf k}}
\newcommand{\vq} {{\mathbf q}}
\newcommand{\vB} {{\mathbf B}}
\newcommand{\bfsigma}{\mbox{\boldmath$\sigma$}}
\newcommand{\nn} {\nonumber}

\begin{document}

\title{Overhauser's spin-density wave in exact-exchange spin density 
functional theory} 
 
\author{S. Kurth}  
\affiliation{Nano-Bio Spectroscopy Group, 
Dpto. de F\'{i}sica de Materiales, 
Universidad del Pa\'{i}s Vasco UPV/EHU, Centro F\'{i}sica de Materiales 
CSIC-UPV/EHU, Av. Tolosa 72, E-20018 San Sebasti\'{a}n, Spain} 
\affiliation{IKERBASQUE, Basque Foundation for Science, E-48011 Bilbao, Spain}
\affiliation{European Theoretical Spectroscopy Facility (ETSF)}
\author{F. G. Eich}   
\affiliation{Institut f\"ur Theoretische Physik, Freie Universit\"at Berlin,   
Arnimallee 14, D-14195 Berlin, Germany}  
\affiliation{Fritz Haber Institute of the Max Planck Society, Faradayweg 4-6, 
D-14195 Berlin, Germany}  
\affiliation{European Theoretical Spectroscopy Facility (ETSF)}

\date{\today}  

\begin{abstract}
The spin density wave (SDW) state of the uniform electron gas is investigated 
in the exact exchange approximation of noncollinear spin density functional 
theory (DFT). Unlike in Hartree-Fock theory, where the uniform paramagnetic 
state of the electron gas is unstable against formation of the spin density 
wave for all densities, in exact-exchange spin-DFT this instability occurs 
only for densities lower than a critical value. It is also shown that, although 
in a suitable density range it is possible to find a non-interacting SDW 
ground state Slater determinant with energy lower than the corresponding 
paramagnetic state, this Slater determinant is not a self-consistent solution 
of the Optimized Effective Potential (OEP) integral equations of noncollinear 
spin-DFT. A selfconsistent solution of the OEP equations which gives an even 
lower energy can be found using an excited-state  Slater determinant where only 
orbitals with single-particle energies in the lower of two bands are occupied 
while orbitals in the second band remain unoccupied even if their 
energies are below the Fermi energy.

\end{abstract}
  
  
\maketitle  
  
\section{Introduction}

The ab-initio description of non-collinear magnetic phenomena 
such as spin-density waves (SDWs) is typically based on an extension 
\cite{BarthHedin:72} of the original Kohn-Sham density functional theory (DFT) 
\cite{KohnSham:65}. As always in DFT, its success crucially relies on the 
accuracy of approximations for the exchange-correlation energy. An important 
step for the application of non-collinear DFT to real systems proved to be 
the construction of a non-collinear version of the local spin density 
approximation (LSDA) by K\"ubler and coworkers 
\cite{KueblerHoeckStichtWilliams:88}. 
An alternative density functional formalism for the description of SDWs and 
antiferromagnetism was proposed by Capelle and Oliveira 
\cite{CapelleOliveira:00,CapelleOliveira:00-2}. In their work the system is 
not described in terms of its density and magnetization density as in usual 
spin-DFT (SDFT) but instead in terms of the density and the so-called 
``staggered density'' where the latter is a nonlocal quantity introduced to 
capture the nonlocal physics of SDWs and antiferromagnetic systems. 

However, nonlocality may also be captured within the framework of usual SDFT 
if one abandons the local approximation to the exchange-correlation energy. 
Orbital functionals, i.e., functionals which explicitly depend on the 
single-particle orbitals rather than on the density (or densities), can be 
highly nonlocal. In DFT (or SDFT), the Optimized Effective Potential (OEP) 
method \cite{TalmanShadwick:76,GraboKreibichKurthGross:00,KuemmelKronik:08} 
provides a framework to treat orbital functionals. This methodology has 
recently been generalized to the case of noncollinear magnetism 
\cite{SharmaDewhurstDraxlKurthHelbigPittalisShallcrossNordstroemGross:07}. 

In the present work we use the noncollinear OEP method to study a very 
simple model system, the uniform electron gas. This model is of paramount 
importance in many-body physics \cite{GiulianiVignale:05}. Moreover, in his 
seminal work Overhauser \cite{Overhauser:60,Overhauser:62} showed analytically 
that this simple model, if treated within the Hartree-Fock (HF) approximation, 
leads to an instability of the paramagnetic phase with respect to 
formation of a spin-density wave. In SDFT, the approximation analogous  to 
HF is the exact-exchange (EXX) approximation which is the HF total energy 
functional but evaluated with orbitals which come from both a {\em local} 
single-particle potential as well as a {\em local} magnetic field. Here we 
use the EXX approximation for a numerical investigation of Overhauser's 
SDW state in the framework of SDFT. This is complementary to another work 
\cite{EichKurthProettoSharmaGross} where Overhauser's SDW state is 
investigated numerically within HF and reduced density-matrix functional 
theory. 

The paper is organized as follows: in Section \ref{sdftnc} we briefly 
review the formalism of noncollinear SDFT and the corresponding OEP method. In 
Section \ref{egas-energy} we minimize the EXX total energy for a given ansatz 
of the SDW state. In Section \ref{selfcons} we investigate if the chosen 
ansatz is selfconsistent in the framework of noncollinear SDFT before 
we provide our conclusions in Section \ref{conclus}. 

\section{Non-collinear spin density functional theory}
\label{sdftnc}

In non-collinear SDFT, a system of interacting electrons with ground state 
$|\Psi_0 \rangle$ moving in an external electrostatic potential $v_{0}(\vr)$ 
(typically the electrostatic potential due to the nuclei) and magnetic 
field $\vB_{0}(\vr)$ is described through its particle density  
\be
n(\vr) = \langle \Psi_0 |\hat{\Psi}^{\dagger}(\vr) \hat{\Psi}(\vr) | \Psi_0 
\rangle
\ee
and its magnetization density
\be
\vm(\vr) = - \mu_B \langle \Psi_0 |\hat{\Psi}^{\dagger}(\vr) \bfsigma 
\hat{\Psi}(\vr) | \Psi_0 \rangle \, .
\ee
Here, $\hat{\Psi}(\vr)$ is the field operator for Pauli spinors, $\mu_B$ is 
the Bohr magneton and $\bfsigma$ is the vector of Pauli matrices (atomic 
units are used throughout). For given external potentials, the total ground 
state energy of such a system can be written as a functional of these two 
densities,
\bea
\lefteqn{
E_{v_{0}, \vB_{0}}[n,\vm] = T_s[n,\vm] + \Id{r} \, v_{0}(\vr) 
n(\vr)}\nn\\ 
&& - \Id{r} \, \vB_{0}(\vr) \vm(\vr) + U[n] + E_{xc}[n,\vm] + E_{ion} 
\label{etot}
\eea 
where $T_s[n,\vm]$ is the non-interacting kinetic energy, 
\be
U[n] = \frac{1}{2} \Id{r} \Id{r'} \; \frac{n(\vr) n(\vr')}{|\vr - \vr'|}
\ee
is the classical electrostatic energy of the electrons and $E_{ion}$ is 
the classical electrostatic energy of the ions. $E_{xc}[n,\vm]$ 
is the (unknown) exchange-correlation energy functional which has to be 
approximated in practice. Once an approximation for this functional is 
specified, the densities $n(\vr)$ and $\vm(\vr)$ can be obtained as 
(ground state) densities of a non-interacting system whose orbitals are 
given by self-consistent solution of the Kohn-Sham (KS) equation 
\be
\left( - \frac{\nabla^2}{2} + v_s(\vr) + \mu_B \bfsigma \vB_s(\vr) \right) 
\Phi_i(\vr) = \varepsilon_i \Phi_i(\vr)
\label{ks-eq}
\ee
where the $\Phi_i(\vr)$ are single-particle Pauli spinors. The effective 
potentials are given by 
\be
v_s(\vr) = v_0(\vr) + \Id{r'} \frac{n(\vr')}{|\vr - \vr'|} + v_{xc}(\vr)
\ee
and
\be
\vB_s(\vr) = \vB_0(\vr) + \vB_{xc}(\vr)
\ee
with the exchange-correlation potentials 
\be
v_{xc}(\vr) = \frac{\delta E_{xc}[n,\vm]}{\delta n(\vr)}
\label{v_xc}
\ee
and
\be
\vB_{xc}(\vr) = - \frac{\delta E_{xc}[n,\vm]}{\delta \vm(\vr)} \; , 
\label{b_xc}
\ee
respectively. 

In this work we will use the EXX energy functional as an approximation to 
the exchange-correlation energy which can be expressed in terms 
of the single-particle spinors $\Phi_i$ as 
\bea
\lefteqn{E_{EXX}[\{\Phi_l \}] = - \frac{1}{2} \sum_{i,j}^{occ} }\nn\\
&& \Id{r} \Id{r'} 
\frac{[\Phi_i^{\dagger}(\vr) \cdot \Phi_j(\vr)][\Phi_j^{\dagger}(\vr') \cdot 
\Phi_i(\vr')]}{|\vr - \vr'|}
\label{exx}
\eea
Since the EXX functional explicitly depends on the (spinor) orbitals but only 
implicitly on the densities, the calculation of the exchange-correlation 
potentials (\ref{v_xc}) and (\ref{b_xc}) has to be performed by means of the 
OEP method 
\cite{TalmanShadwick:76,GraboKreibichKurthGross:00,KuemmelKronik:08}. A 
formulation of this technique for the case of non-collinear magnetism has 
recently been given in Ref.~\onlinecite{SharmaDewhurstDraxlKurthHelbigPittalisShallcrossNordstroemGross:07}. 
The coupled OEP integral equations for the exchange-correlation potentials 
can be obtained by applying the chain rule of functional derivatives
\bea
\frac{\delta E_{xc}}{\delta v_s(\vr)} &=& \Id{r}' \left( v_{xc}(\vr')
\frac{\delta n(\vr')}{\delta v_s(\vr)} - \vB_{xc}(\vr')
\frac{\delta \vm(\vr')}{\delta v_s(\vr)} \right) \nn \\
&=& \sum_{i}^{occ} \Id{r}' \left( \frac{\delta E_{xc}}{\delta \Phi_i(\vr')}
\frac{\delta \Phi_i(\vr')}{\delta v_s(\vr)} + h.c.
\right)
\eea
\bea
\frac{\delta E_{xc}}{\delta \vB_s(\vr)} &=& \Id{r}' \left( v_{xc}(\vr')
\frac{\delta n(\vr')}{\delta \vB_s(\vr)} - \vB_{xc}(\vr')
\frac{\delta \vm(\vr')}{\delta \vB_s(\vr)} \right) \nn \\
&=& \sum_{i}^{occ} \Id{r}' \left( \frac{\delta E_{xc}}{\delta \Phi_i(\vr')}
\frac{\delta \Phi_i(\vr')}{\delta \vB_s(\vr)} + h.c.
\right)
\eea
The functional derivatives of the spinor orbitals and the densities with 
respect to the potentials can be computed from first-order perturbation theory 
and, following the notation of Ref.~\onlinecite{KurthPittalis:06}, the 
OEP equations can be written in compact form as 
\be
\sum_{i}^{occ} \left( \Phi_i^{\dagger}(\vr) \Psi_i(\vr) + h.c. \right) = 0
\label{oep-s1}
\ee
\be
- \mu_B \sum_{i}^{occ} \left( \Phi_i^{\dagger}(\vr) \bfsigma \Psi_i(\vr) +
h.c. \right) = 0
\label{oep-s2}
\ee
where we have defined the orbital shifts
\cite{GraboKreibichKurthGross:00,KuemmelPerdew:03,KuemmelPerdew:03-2}
\be
\Psi_i(\vr) = \sum_{\stackrel{j}{j \neq i}}
\frac{D_{ij} \Phi_j(\vr)}{\varepsilon_i - \varepsilon_j}
\label{orb-shifts}
\ee
with 
\bea
\lefteqn{
D_{ij} = \Id{r}' \, \Phi_{j}^{\dagger}(\vr') }\nn\\
&&\left( \left( v_{xc}(\vr') +
\mu_B \bfsigma \vB_{xc}(\vr') \right) \Phi_i(\vr') -
\frac{\delta E_{xc}}{\delta \Phi_i^{\dagger}(\vr') } \right)\;\; .
\eea
Eqs.~(\ref{ks-eq}) - (\ref{exx}) and (\ref{oep-s1}) - (\ref{oep-s2}) 
constitute our formal framework to investigate the spin density wave 
in the uniform electron gas in exact exchange. 

\section{Uniform electron gas with Spin-Density Wave: direct minimization of 
energy}
\label{egas-energy}

We study a uniform electron gas, i.e., a system of electrons with spatially 
constant density moving in the electrostatic potential created by a 
neutralizing, uniform density of positive background charge. 

In the following, rather than calculating the Kohn-Sham potentials 
selfconsistently, we {\em assume} that the Kohn-Sham 
electrostatic potential $v_s(\vr)$ is a constant (which we set to zero) and 
that the Kohn-Sham magnetic field $\vB_s(\vr)$ forms a spiral with amplitude 
$B$ and wavevector $\vq = q \ve_z$ where $\ve_z$ is a unit vector in 
$z$-direction, i.e.,  $\vB_s(\vr) = (B \cos(q z), B \sin(q z), 0)$. 
Moreover, we also assume that the Kohn-Sham magnetic field is entirely due to 
its exchange-correlation part, i.e., we study the system without external 
magnetic field. Of course, at some point we have to verify that our 
assumptions are consistent within the SDFT framework. This question will be 
studied in Section \ref{selfcons}.

With the effective potentials given above, the Kohn-Sham equation 
(\ref{ks-eq}) can be solved analytically. 
A complete set of quantum numbers is given by a
band index $b=1,2$ and a wavevector $\vk$. The corresponding single-particle
eigenstates and eigenenergies for the first band are given by
\be
\Phi_{\vk}^{(1)}(\vr) = \frac{\exp(i \vk \vr)}{\sqrt{\cal{V}}} \left(
\begin{array}{c}
\cos \vartheta_{k_z} \\
\sin \vartheta_{k_z} \exp(i q z)
\end{array} \right)
\label{orb-1}
\ee
where $\cal{V}$ is the system volume (which tends to infinity) and
\be
\varepsilon_{\vk}^{(1)}  = \frac{k_{\|}^2}{2} + \varepsilon_{\kappa}^{(1)}
\label{band-1}
\ee
where $k_{\|}= \sqrt{k_x^2 + k_y^2}$, $\kappa = k_z + \frac{q}{2}$ and 
\be
\varepsilon_{\kappa}^{(1)} = 
\frac{\kappa^2}{2} + \frac{q^2}{8} - \sqrt{\frac{q^2}{4} \kappa^2 +
\mu_B^2 B^2} \, .
\ee
The angle $\vartheta_{k_z}$ is defined through the relation
\be
\tan 2 \vartheta_{\kappa} = -2 \alpha
\label{angle-1}
\ee
with
\be
\alpha = \frac{\mu_B B}{q \kappa} \; .
\ee
For the second band the eigenstates and eigenenergies are
\be
\Phi_{\vk}^{(2)}(\vr) = \frac{\exp(i \vk \vr)}{\sqrt{\cal{V}}} \left(
\begin{array}{c}
-\sin \vartheta_{k_z} \\
\cos \vartheta_{k_z} \exp(i q z)
\end{array} \right)
\label{orb-2}
\ee
and
\be
\varepsilon_{\vk}^{(2)}  = \frac{k_{\|}^2}{2} + \varepsilon_{\kappa}^{(2)}
\label{band-2}
\ee
with
\be
\varepsilon_{\kappa}^{(2)} = 
\frac{\kappa^2}{2} + \frac{q^2}{8} + \sqrt{\frac{q^2}{4} \kappa^2 +
\mu_B^2 B^2} \, .
\ee
In order for the definition of $\Phi_{\vk}^{(1)}(\vr)$ and 
$\Phi_{\vk}^{(2)}(\vr)$ to be unique, the angle $\vartheta_{\kappa}$ has to be 
restricted to an interval of length $\pi/2$. Assuming that $B \geq 0$ and 
$q>0$, we find from Eq.~(\ref{angle-1}) that $-\pi/2 < \vartheta_{\kappa} 
\leq 0$. Using a trigonometric identity we can transform Eq.~(\ref{angle-1}) to
\be
\tan \vartheta_{\kappa} = \frac{1}{2 \alpha} \left( 1 - \sqrt{1+ 4 \alpha^2} 
\right)
\label{angle-2}
\ee
From Eq.~(\ref{angle-1}) we see that for finite $B$ and $\kappa = 0$ the 
angle $\vartheta_{\kappa=0}=-\frac{\pi}{4}$. In order for $\vartheta_{\kappa}$ 
to be a continuous function of $\kappa$ with values in the correct range we 
invert Eq.~(\ref{angle-2}) as
\be
\vartheta_{\kappa} = \left\{
\begin{array}{cc}
\arctan\left( \frac{1}{2 \alpha} \left( 1 - \sqrt{1+ 4 \alpha^2} \right)
\right) & \mbox{  for $\kappa \geq 0 $} \\
- \frac{\pi}{2} + \arctan\left( \frac{1}{2 \alpha}
\left( 1 - \sqrt{1+ 4 \alpha^2} \right) \right) &
\mbox{  for $\kappa < 0$}
\end{array}
\right.
\label{angle-3}
\ee
With the single-particle states fully defined we can write down the 
uniform electronic (ground state) density as 
\bea
n &=& \sum_b n^{(b)} = 
\sum_b \ID{k} \; \theta(\varepsilon_F - \varepsilon_{\vk}^{(b)}) \nn \\
&=& \frac{1}{4 \pi^2} \sum_b \int {\mathrm d}{\kappa} \, \theta(\varepsilon_F - 
\varepsilon_{\kappa}^{(b)}) (\varepsilon_F - \varepsilon_{\kappa}^{(b)})
\label{dens}
\eea
where $n^{(b)}$ is the density contribution of band $b$, $\theta(x)$ is the 
Heaviside step function and the trivial integrals have been carried out in 
the last step. Using Eqs.~(\ref{band-1}) and (\ref{band-2}), the integration 
limits can be determined analytically and the remaining integral can easily 
be solved in closed form but we refrain from giving the explicit expression 
here. 

Similarly, we can compute the magnetization density. We obtain for the $x-$ 
and $y-$ components 
\be
m_x(\vr) = m_0 \cos(q z) 
\ee
and 
\be
m_y(\vr) = m_0 \sin(q z) 
\ee
where the amplitude of the spin-density wave is 
\bea
m_0 &=& - \frac{\mu_B}{2 \pi^2} \sum_b {\rm sign}(b) \nn\\ 
&&\int {\mathrm d}{\kappa} \,\;
\theta(\varepsilon_F - \varepsilon_{\kappa}^{(b)})\; (\varepsilon_F - 
\varepsilon_{\kappa}^{(b)}) \; \sin\vartheta_{\kappa} \cos\vartheta_{\kappa} 
\;\;\;
\label{sdw-amp}
\eea
and we have defined
\be
{\rm sign}(b) = \left\{ 
\begin{array}{cl} 
+1 & \mbox{ for $b=1$} \\
-1 & \mbox{ for $b=2$}
\end{array}
\right. \; .
\ee
Using the symmetry relation $\theta_{-\kappa} = - \frac{\pi}{2} - 
\theta_{\kappa}$ (see Eq.~(\ref{angle-3})), the $z-$component of the 
magnetization density can be shown to vanish identically, 
\be
m_z(\vr) = 0 \; \; .
\ee
We point out that the vector of the magnetization density here is 
parallel to the Kohn-Sham magnetic field. This certainly is a 
consequence of the simplicity of the system under study here. For more 
complicated systems it was shown in 
Ref.~\onlinecite{SharmaDewhurstDraxlKurthHelbigPittalisShallcrossNordstroemGross:07} 
that these quantities need not be parallel in noncollinear SDFT in EXX. 
This is an important difference to the noncollinear LSDA formulation 
of Ref.~\onlinecite{KueblerHoeckStichtWilliams:88} where the magnetization 
density and the exchange-correlation magnetic field are locally parallel by 
construction. 

We now turn to the evaluation of the energy of Eq.~(\ref{etot}) and note that 
for an electrically neutral system with uniform ionic and electronic densities 
the sum of the ionic energy $E_{ion}$, the 
electronic interaction with the ionic potential $\Id{r}\; v_0(\vr) n(\vr)$, and 
the Hartree energy $U[n]$ exactly cancels out. We study the system at 
vanishing external magnetic field, $\vB_0(\vr) \equiv 0$, and use the exact 
exchange energy of Eq.~(\ref{exx}) as an approximation to the 
exchange-correlation energy functional. It is expected 
\cite{GiulianiVignale:05,GiulianiVignale:08} that inclusion of 
correlation leads to SDW states higher in energy than the paramagnetic 
states.

In our case the total energy per unit volume only consists of a kinetic 
and an exchange contribution, i.e., 
\be
\tilde{e}_{tot} = \tilde{t}_s + \tilde{e}_{EXX} \; .
\ee
The kinetic energy per unit volume has contributions from the two bands, 
\be
\tilde{t}_s = \frac{T_s}{\cal{V}} = \sum_b \tilde{t}_s^{(b)} \, ,
\ee
where the contribution of the first band  is given by
\bea
\tilde{t}_s^{(1)} &=&
\frac{1}{8 \pi^2} \int {\mathrm d}{\kappa} \, \theta(\varepsilon_F - 
\varepsilon_{\kappa}^{(1)}) \left( \varepsilon_F - \varepsilon_{\kappa}^{(1)} 
\right) \nn \\
&& \left( \varepsilon_F - \varepsilon_{\kappa}^{(1)} + \kappa^2 + 
\frac{q^2}{4} + 2 q \kappa \, \sin^2 \vartheta_{\kappa} \right)
\label{tkin-1}
\eea
while the contribution of the second band is
\bea
\tilde{t}_s^{(2)} &=&
\frac{1}{8 \pi^2} \int {\mathrm d}{\kappa} \, \theta(\varepsilon_F - 
\varepsilon_{\kappa}^{(2)}) \left( \varepsilon_F - \varepsilon_{\kappa}^{(2)} 
\right) \nn \\
&& \left( \varepsilon_F - \varepsilon_{\kappa}^{(2)} + \kappa^2 + 
\frac{q^2}{4} + 2 q \kappa \, \cos^2 \vartheta_{\kappa} \right) \; .
\label{tkin-2}
\eea
Inserting the orbitals (\ref{orb-1}) and (\ref{orb-2}) into Eq.~(\ref{exx}), 
the exchange energy per unit volume, $\tilde{e}_{EXX}$, can also be expressed 
as sum of two terms, 
\be
\tilde{e}_{EXX} = \tilde{e}_{EXX}^{(1)} + \tilde{e}_{EXX}^{(2)} \; .
\ee
The first term which describes intraband exchange is, after carrying out the 
angular integrals, given by 
\bea
\lefteqn{
\tilde{e}_{EXX}^{(1)} = - \frac{1}{32 \pi^3}  }\nn \\
&& \sum_b \int {\mathrm d}{\kappa} \, 
\theta(\varepsilon_F - \varepsilon_{\kappa}^{(b)})  
\int {\mathrm d}{\kappa'} \theta(\varepsilon_F - \varepsilon_{\kappa'}^{(b)}) 
\nn \\
&& \cos^2(\vartheta_{\kappa} - \vartheta_{\kappa'}) 
I(y^{(b)}(\kappa),y^{(b)}(\kappa'),(\kappa-\kappa')^2) 
\label{exx-1}
\eea
while the second term, describing interband exchange, reads
\bea
\lefteqn{
\tilde{e}_{EXX}^{(2)} = - \frac{1}{16 \pi^3}  }\nn \\
&& \int {\mathrm d}{\kappa} \, 
\theta(\varepsilon_F - \varepsilon_{\kappa}^{(1)})  
\int {\mathrm d}{\kappa'} \theta(\varepsilon_F - \varepsilon_{\kappa'}^{(2)}) 
\nn \\
&& \sin^2(\vartheta_{\kappa} - \vartheta_{\kappa'}) 
I(y^{(1)}(\kappa),y^{(2)}(\kappa'),(\kappa-\kappa')^2) \; .
\label{exx-2}
\eea
where we have defined 
\be
y^{(b)}(\kappa) = 2 ( \varepsilon_F - \varepsilon_{\kappa}^{(b)}) \; .
\ee
In Eqs.~(\ref{exx-1}) and (\ref{exx-2}) we also have used the integral
\bea
\lefteqn{
I(y_1,y_2,a) =}\nn \\ 
&& \int_0^{y_1} {\mathrm d}{y} \, \int_0^{y_2} {\mathrm d}{y'} \, 
\frac{1}{\sqrt{(y-y')^2 + 2(y+y') a + a^2}} \;\;\;\;
\label{exx_integral}
\eea
which can be solved in closed form by transforming to new integration variables 
$z=y-y'$ and $z'=(y+y')/2$ and changing the integration limits accordingly. 
Therefore the calculation of the total energy only requires the numerical 
calculation of a two-dimensional integral. 

We have calculated the total energy per particle
\be
e_{tot} = \frac{\tilde{e}_{tot}}{n}
\ee
in  the following way: we start by numerically calculating 
the Fermi energy for a given value $n$ of the density or, equivalently, the 
Wigner-Seitz radius 
\be
r_s = \left( \frac{3}{4 \pi n} \right)^{1/3} \; ,
\ee
and given values of the parameters $B$ and $q$ from Eq.~(\ref{dens}). The 
Fermi energy thus becomes a function of these three parameters,
\be 
\varepsilon_F=\varepsilon_F(r_s,q,B) \; ,
\label{efermi}
\ee 
which is then used to evaluate the total energy per particle for these 
parameter values. We then have, for fixed $r_s$, minimized $e_{tot}$ as a 
function of the parameters $q$ and $B$ numerically. 

\begin{figure}[tb]
\includegraphics[width=.47\textwidth]{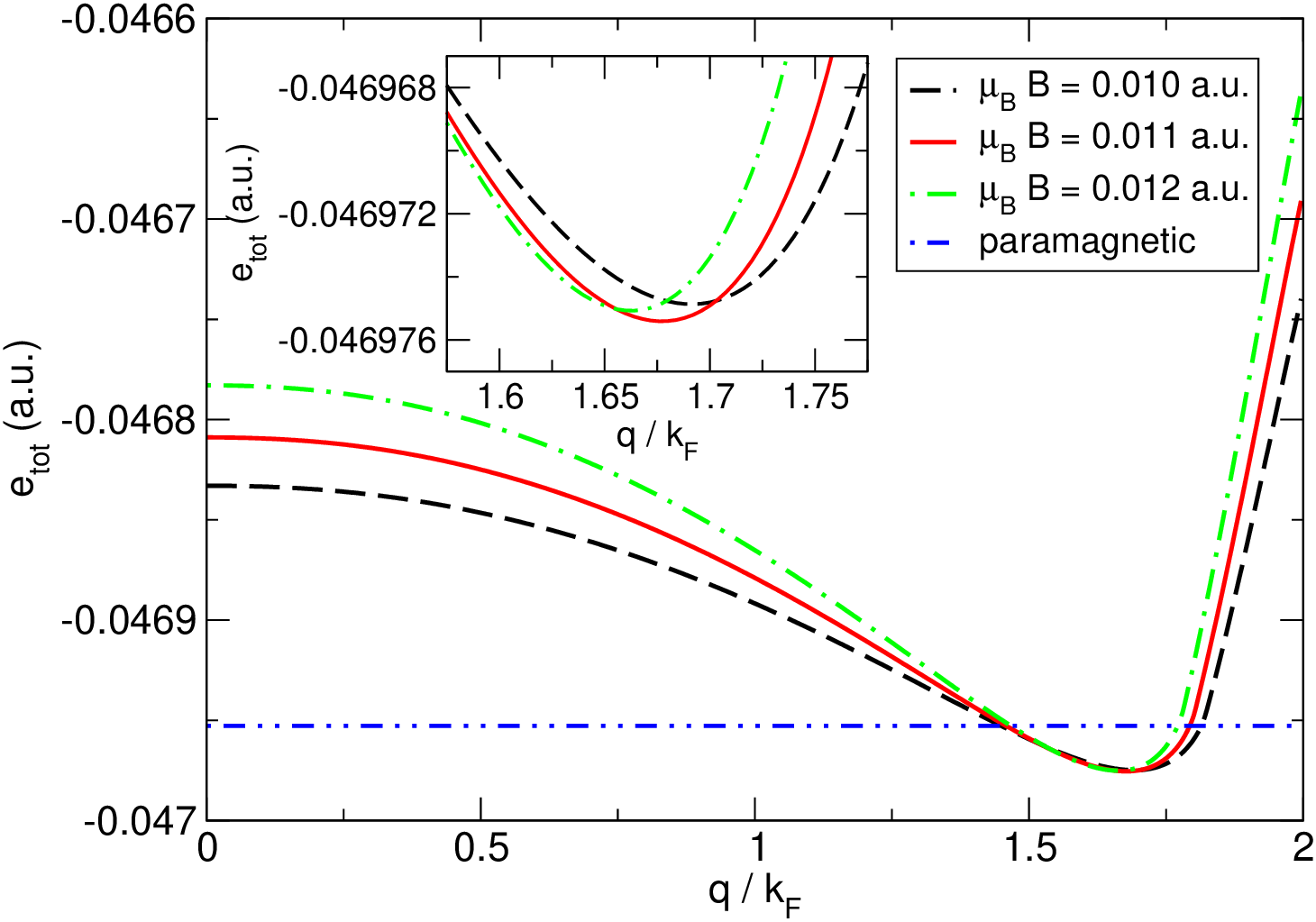}
\caption{Total energy per particle in EXX for the electron gas at $r_s=5.4$ 
with spin density wave as function of $q/k_F$ and different values of the 
amplitude $B$ of the Kohn-Sham magnetic field. The straight line corresponds 
to the total energy per particle of the paramagnetic state at this density. 
The inset shows a magnification close to the minimum.}
\label{toten_rs5p4}
\end{figure}

In Fig.~\ref{toten_rs5p4} we show the total energy per electron at $r_s=5.4$ 
for a few values of $B$ as function of $q/k_F$ where 
\be
k_F = \left( \frac{9 \pi}{4} \right)^{1/3} \frac{1}{r_s}
\ee
is the Fermi wavenumber of the uniform electron gas in the paramagnetic state. 
The value $r_s=5.4$ was chosen because then i) the SDW phase is lower in energy 
than both the paramagnetic and ferromagnetic phases and ii) the 
amplitude of the SDW (or the KS magnetic field) is relatively high 
such that the resulting energy differences can easily be resolved numerically. 
We clearly see that for the given values of $B$ for wavenumbers between 
$q/k_F \approx 1.5$ and $q/k_F \approx 1.75$ the energy of the SDW state is 
lower than the energy of the paramagnetic state. The lowest energy for this 
value of $r_s$ is achieved for the parameters $\mu_B B = 0.011$ a.u. and 
$q/k_F=1.68$. 

\begin{figure}[tb]
\includegraphics[width=.47\textwidth]{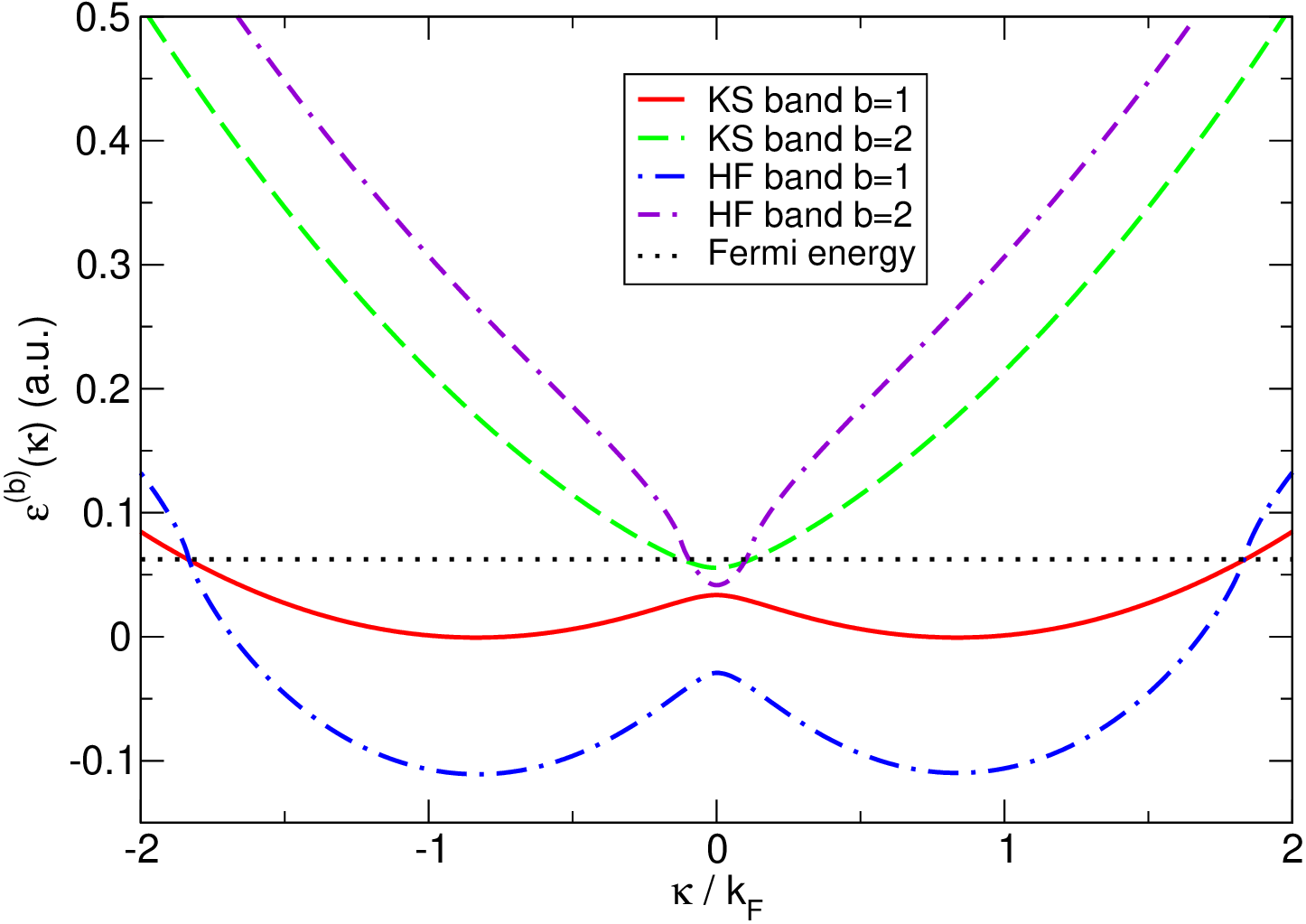}
\caption{Single-particle KS and HF bands at $k_{\|}=0$ for the optimized 
parameter values 
($\mu_B B = 0.011$ a.u. and $q/k_F=1.68$ at $r_s=5.4$) mimizing the EXX total 
energy per particle in Fig.~\protect\ref{toten_rs5p4}. The straight line 
indicates the Fermi energy and shows that close to $\kappa/k_F=0$ states of 
the second KS band [dashed (green) line] are occupied in the ground state. The 
KS orbitals have been used to compute the HF Hamiltonian and the resulting 
HF bands have been shifted rigidly such that the lower HF and KS bands 
equal $\varepsilon_F$ at the same value of $|\kappa/k_F|$. The relative 
position of the second HF band [dash-dash-dotted (purple) line] indicates that 
also in HF states in both bands will be occupied.}
\label{bands_rs5p4}
\end{figure}

In Fig.~\ref{bands_rs5p4} we show the KS single-particle 
dispersions of Eqs.~(\ref{band-1}) and (\ref{band-2}) as well as the HF 
single particle dispersions. To obtain the latter ones we first calculate 
the HF self energy (which is a $2 \times 2$ matrix in spin space) as 
\be
\Sigma^{HF}(\vr,\vr') = - \sum_b \sum_{\vk} \theta(\varepsilon_F - 
\varepsilon_{\vk}^{(b)})  \frac{\Phi_{\vk}^{(b)}(\vr) \otimes 
\Phi_{\vk}^{(b) \dagger}(\vr') }{|\vr - \vr'|}
\label{sigma_hf}
\ee
and then diagonalize the resulting HF Hamiltonian
\be 
\hat{h}^{HF} = - \frac{\nabla^2}{2} + \Id{r'} \; \Sigma^{HF}(\vr,\vr') 
\ldots
\ee
where the second term is a to be read as an integral operator. We would 
like to emphasize that we use the KS orbitals and orbital energies to 
evaluate the HF self energy, i.e., we do {\em not} perform a selfconsistent 
HF calculation here. 

In Fig.~\ref{bands_rs5p4} we show the KS and HF dispersions only for the 
$\kappa$-coordinate, i.e., we set $k_{\|}=0$. As expected, close to 
$\kappa/k_F=0$ a direct gap opens up in the KS single-particle dispersions due 
to the presence of the spin-density wave. The position of the Fermi energy is 
such that not only states of the lower ($b=1$) band but also states of the 
second ($b=2$) KS band are occupied in the ground state. 

The HF bands in Fig.~\ref{bands_rs5p4} have been rigidly shifted by a constant 
such that the lower HF band ($b=1$) and the lower KS band equal the Fermi 
energy for the {\em same} value of $\kappa/k_F$. It is evident that, as 
expected, the HF single particle direct band gap at $\kappa/k_F=0$ is much 
larger than the corresponding KS gap. Moreover, the position of the second 
HF band indicates that also in the HF case there will be occupied states 
in the second band. While here we have calculated the HF bands using 
the DFT orbitals and orbital energies, we have confirmed 
\cite{EichKurthProettoSharmaGross} that the above statement is true also for 
a HF energy minimization and the resulting HF bands are very close to the 
ones presented here.

The occupation of states in both single-particle bands is sometimes excluded 
in works on the SDW in the Hartree-Fock approximation 
\cite{Overhauser:62,GiulianiVignale:05} and also in a numerical investigation 
\cite{EichKurthProettoSharmaGross} we have found that for the global energy 
minimum in Hartree-Fock only 
the lowest single-particle band is occupied. This has motivated us to do the 
minimization of the total energy in EXX also under the additional constraint 
that only states of the lowest subband are occupied. 

\begin{figure}[tb]
\includegraphics[width=.47\textwidth]{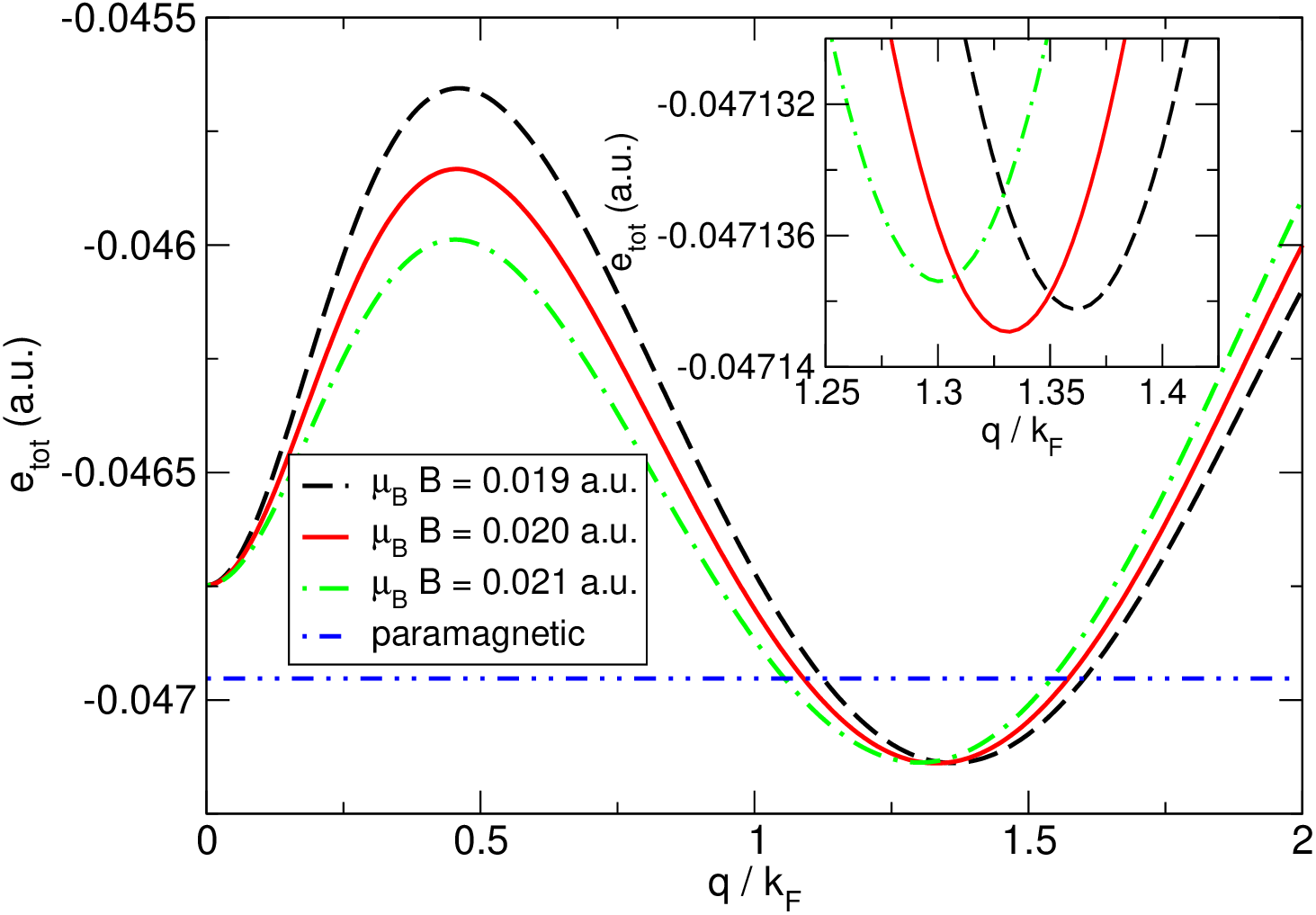}
\caption{Same as Fig.~\protect\ref{toten_rs5p4} except that now only states 
in the lower band are allowed to be occupied. The total energy per 
particle at the minimum is lower than when states in both bands are allowed 
to be occupied.}
\label{toten_rs5p4_one}
\end{figure}

Similar to  
Fig.~\ref{toten_rs5p4}, in Fig.~\ref{toten_rs5p4_one} we show the total energy 
per electron at $r_s=5.4$ for a few values of $B$ as function of $q/k_F$. 
Of course, the constrained minimization leads, for a given value of $r_s$, to 
different optimized parameter values. Surprisingly, however, we found that the 
minimization constraining the occupation to the lower subband {\em leads to 
lower total energies} than the ones obtained with a two-band minimization. 
Moreover, this lower total energy is achieved with a Slater determinant 
{\em which has empty states below the Fermi level.}
 
\begin{figure}[tb]
\includegraphics[width=.47\textwidth]{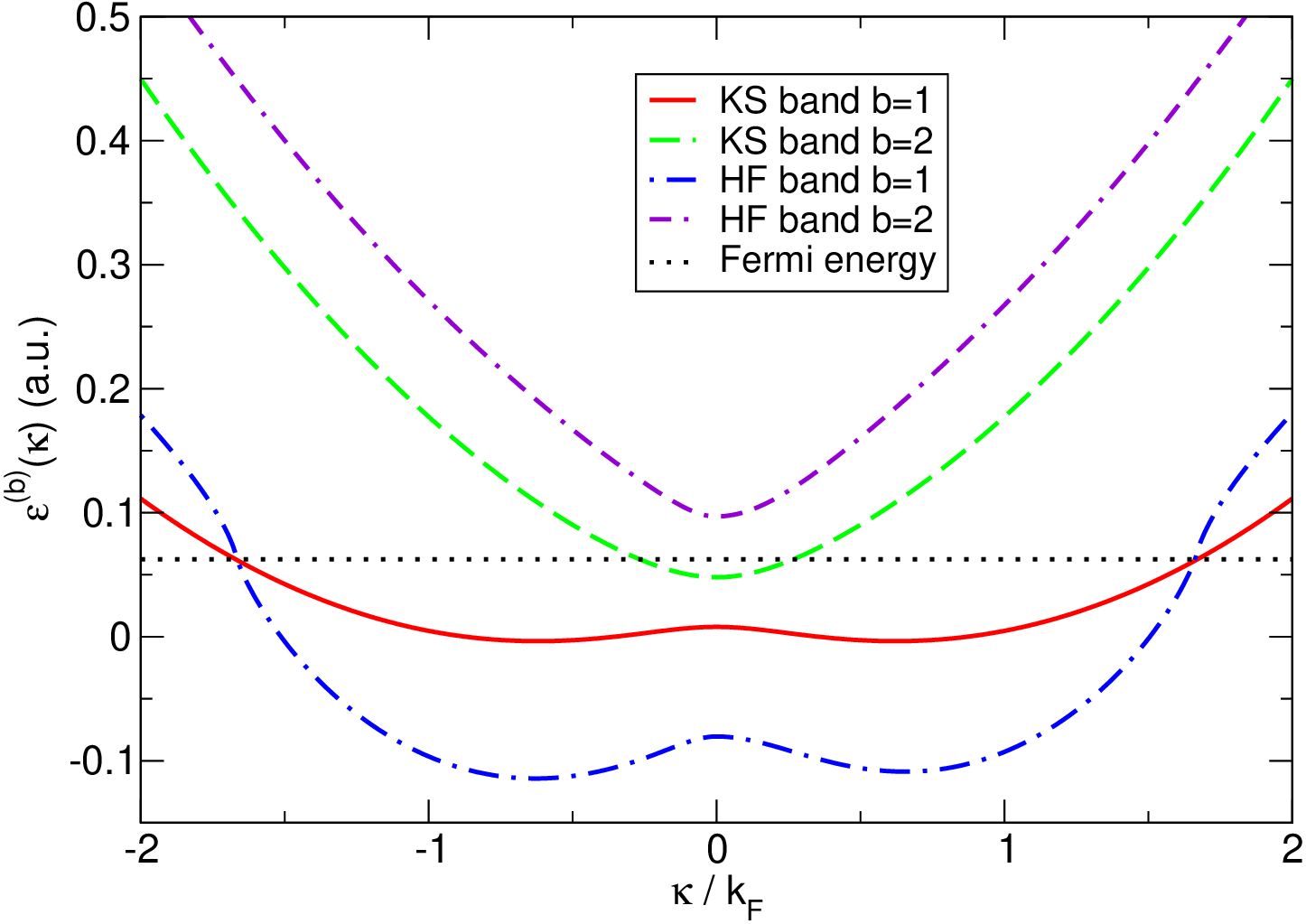}
\caption{Same as Fig.~\protect\ref{bands_rs5p4} except that the optimized 
parameters are used which result from a minimization with occupied states in 
the lower KS band only. For $r_s=5.4$ these values are 
$\mu_B B = 0.020$ a.u. and $q/k_F=1.33$. Again, the straight line indicates 
the Fermi energy. Note that the states of the second KS band [dashed (green) 
line] remain unoccupied 
in this calculation, even if their single-particle energies are below the 
Fermi level, i.e., the resulting Slater determinant is {\em not} a ground 
state of the Kohn-Sham problem. On the other hand, the post-hoc evaluation of 
the HF bands (for details see caption of Fig.~\protect\ref{bands_rs5p4} and 
the main text) indicates that the second HF band [dash-dash-dotted (purple)  
line] will remain unoccupied and the resulting HF wavefunction will be a 
ground state Slater determinant.}
\label{bands_rs5p4_one}
\end{figure}

This can be seen in Fig.~\ref{bands_rs5p4_one} where we show the KS and HF 
energy bands at $r_s=5.4$ for this ``one-band'' minimization for the optimized 
parameter values of $\mu_B B = 0.020$ a.u. and $q/k_F=1.33$. We see that there 
are states in the second KS band below the Fermi energy which, due to 
the constraint in the minimization, remain unoccupied.
We also note that for the one-band case 
the amplitude of the minimizing Kohn-Sham magnetic field, and therefore also 
the ``gap'' between the two KS bands at $\kappa/k_F=0$, is almost twice as 
large as in the two-band case. Compared to Fig.~\ref{bands_rs5p4}, the 
intersection of the Fermi energy with the bands $\varepsilon^{(b)}(\kappa)$ is 
shifted to a lower value of $|\kappa|$. 

The HF bands again have rigidly 
been shifted such that the lower HF and KS bands intersect the Fermi energy 
at the same $\kappa$. Again, the direct HF gap at $\kappa/k_F=0$ is 
significantly larger than the KS gap. In contrast to the two-band case, the 
second HF band now is energetically higher than the Fermi energy and the 
corresponding HF state would, unlike the EXX state, have no unoccupied 
single-particle states below $\varepsilon_F$. Again here we have done only 
a post-hoc evaluation of the HF bands but we have checked that the statement 
remains valid for a selfconsistent HF calculation as well 
\cite{EichKurthProettoSharmaGross}. 

\begin{figure}[tb]
\includegraphics[width=.47\textwidth]{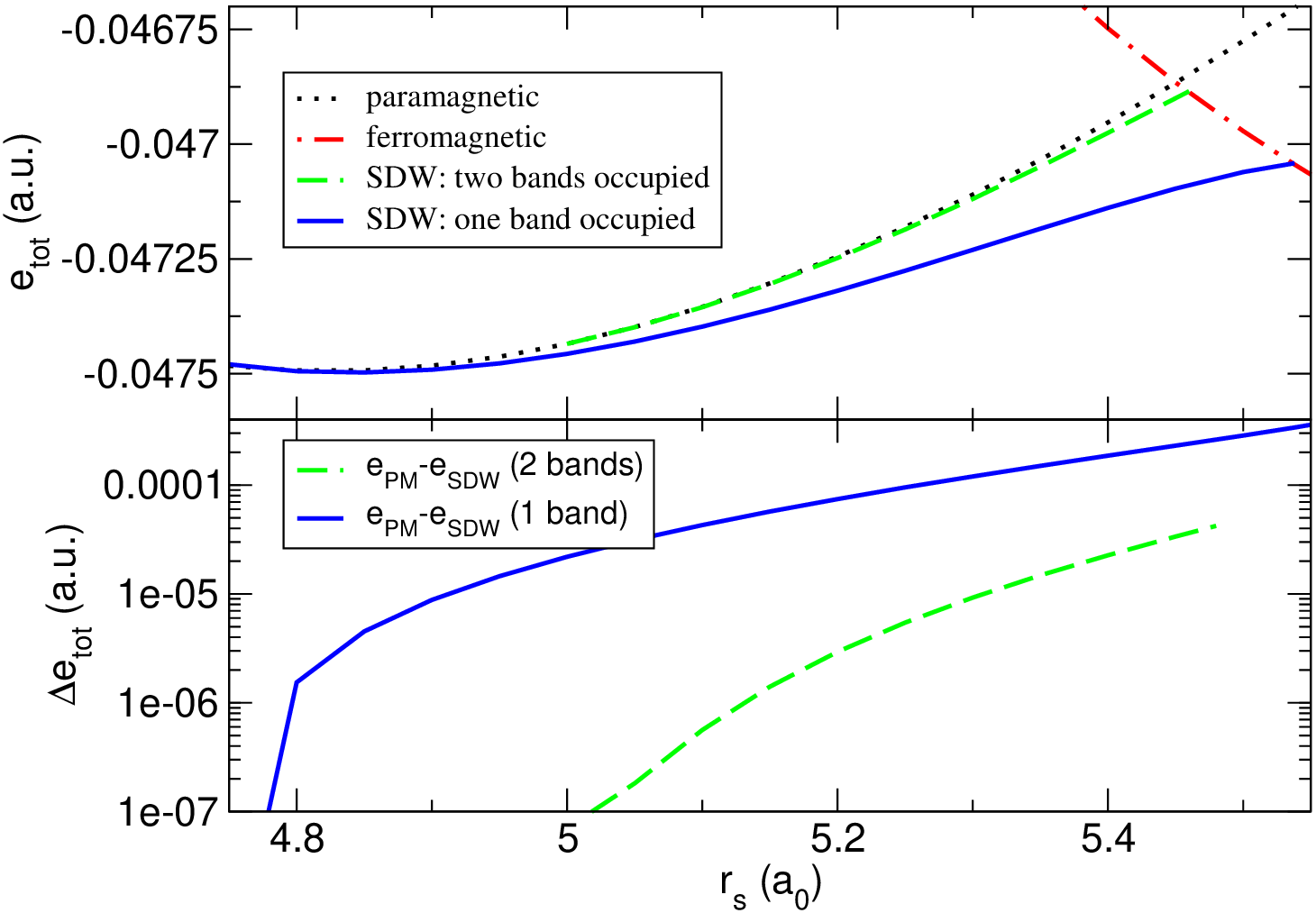}
\caption{Upper panel: Total energy per particle in EXX for different phases of 
the uniform electron gas as function of Wigner-Seitz radius $r_s$. In the SDW 
phase, in one case the occupation of single-particle states in both bands is 
allowed while in the other case the occupied states are restricted to the 
lower band. Lower panel: energy difference between the total energies 
per particle of the paramagnetic phase and the SDW phase for SDWs with 
occupied states in one and two bands. For the two-band case, the 
SDW phase is lower in energy than both the paramagnetic and the 
ferromagnetic phase for $5.0 \stackrel{<}{\sim} r_s \stackrel{<}{\sim} 5.46$. 
For the one-band case the range of stability of the SDW phase is 
$4.78 \stackrel{<}{\sim} r_s \stackrel{<}{\sim} 5.54$.}
\label{phasediag}
\end{figure}

We have optimized the EXX total energy per particle for a range of $r_s$-values 
once for single-particle occupations in both energy bands and once for 
occupations restricted to the lower band. In Fig.~\ref{phasediag} we show the 
resulting phase diagram in the relevant density range. When allowing 
occupations in both bands, the SDW state (which is then a ground state Slater 
determinant) is lower in energy than both the 
paramagnetic and the ferromagnetic phase for $r_s$ in the range $5.0 
\stackrel{<}{\sim} r_s \stackrel{<}{\sim} 5.46$. In this case the energies 
are very close to the energies of the paramagnetic phase (energy differences 
of less than $4\times 10^{-5}$ a.u., see lower panel of Fig.~\ref{phasediag}) 
and therefore the transition to the ferromagnetic phase occurs at an value 
of $r_s$ only slightly higher than the $r_s$-value where paramagnetic and 
ferromagnetic phase are degenerate. 

On the other hand, restricting the single-particle occupation 
to the lowest band, the SDW state is more stable than para- and ferromagnetic 
state for $4.78 \stackrel{<}{\sim} r_s \stackrel{<}{\sim} 5.54$. In this 
case the energy differences between the paramagnetic and the SDW phase 
range to almost $4\times10^{-4}$ a.u. (lower panel of Fig.~\ref{phasediag}), 
almost an order of magnitude larger than in the two-band case. However, for 
all $r_s$ values in the stability range of the SDW phase, the minimizing 
Slater determinant in the one-band case is {\em not} a ground state of the 
Kohn-Sham problem. 

Both the one- and two-band cases in EXX have in common that they predict 
the SDW phase to be lower in energy than the paramagnetic phase only for a  
restricted range of $r_s$ values. This is different from the Hartree-Fock 
case \cite{Overhauser:62,GiulianiVignale:05} where the SDW phase is more 
stable than the paramagnetic phase for {\em all} values of $r_s$. This is not 
completely surprising since due to the additional constraint of local 
Kohn-Sham potentials $v_s$ and $\vB_s$ in the EXX minimization, the resulting 
energies have to be higher than the Hartree-Fock total energies. Since for 
small values of $r_s$ the SDW total energies in HF are extremely close to the 
total energies of the paramagnetic phase \cite{EichKurthProettoSharmaGross}, 
the higher EXX total energies can easily lead to a more stable paramagnetic 
phase. 

\begin{figure}[tb]
\includegraphics[width=.47\textwidth]{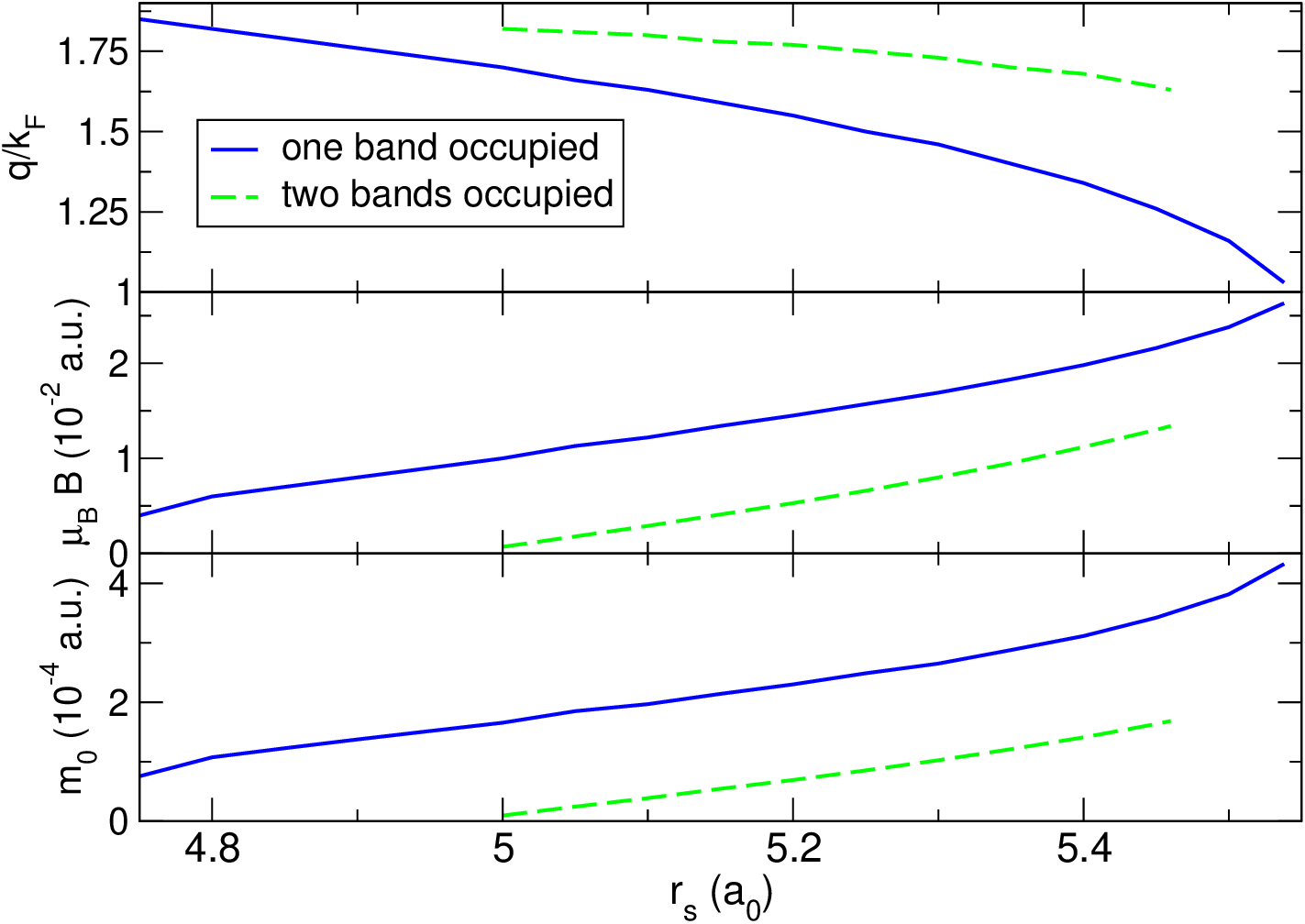}
\caption{Optimized values for the parameters $q$ (upper panel) and $B$ (middle 
panel) for which the EXX total energy per particle of the SDW phase is 
minimized. The results are shown over the range of $r_s$-values for which the 
SDW phase is lower in energy than both the paramagnetic and the ferromagnetic 
phases for the cases when both bands or only one band are occupied. 
Lower panel: amplitude of the SDW (Eq.~(\protect\ref{sdw-amp})) for the one- 
and two-band case.}
\label{opt_param}
\end{figure}

In Fig.~\ref{opt_param} we show the SDW parameters $q$ (upper panel) and $B$ 
(middle panel) for which the EXX total energy per particle is minimized in the 
one- and two-band cases for those $r_s$-values for which the SDW phase is more 
stable than both paramagnetic and ferromagnetic phases. For the one-band case, 
the wavenumber $q$ of the spin-density wave covers almost the whole range 
between $k_F$ and $2 k_F$ while for the two-band case this range is much 
narrower. The amplitudes $B$ and $m_0$ of the Kohn-Sham magnetic field 
(middle panel) and the magnetization density (lower panel) of the SDW are 
significantly smaller in the two-band case as for the case with occupied 
single-particle states in one band only.
It is sometimes assumed \cite{GiulianiVignale:05} that the 
wavenumber of the SDW is close to $2 k_F$. Our results show that this need not 
be the case, as in the one-band case $q$ approaches $k_F$ for densities at 
the lower end of the stability range of the SDW phase. However, neither in EXX 
nor in HF \cite{EichKurthProettoSharmaGross} have we ever found a stable 
SDW state with wavenumber lower than $k_F$. 

\section{Selfconsistency}
\label{selfcons}

In the previous Section we have used an ansatz for the Kohn-Sham orbitals 
in the SDW phase which depends on two parameters and then minimized the EXX 
total energy per particle with respect to these parameters. We have done this 
minimization once allowing single-particle states in both bands to be 
occupied and once for occupations only in the lower band. This is different 
from the usual way of applying DFT where one calculates the 
exchange-correlation potentials and solves the Kohn-Sham equation 
self-consistently. In our case, the calculation of the EXX potentials requires 
solution of the OEP equations (\ref{oep-s1}) and (\ref{oep-s2}). In the 
present Section we still use our ansatz for the Kohn-Sham orbitals and 
investigate if it is consistent with the OEP equations. 

We start by calculating the orbital shifts of Eq.~(\ref{orb-shifts}) in the 
EXX approximation. Inserting our ansatz after some straightforward algebra we 
obtain for the orbital shift of the first band 
\be
\Psi_{\vk}^{(1)}(\vr) = \frac{G(\vk)}{\varepsilon_{\kappa}^{(1)} - 
\varepsilon_{\kappa}^{(2)}} \Phi_{\vk}^{(2)}(\vr)
\label{shift-1}
\ee
while the shift for the second band reads
\be
\Psi_{\vk}^{(2)}(\vr) = - \frac{G(\vk)}{\varepsilon_{\kappa}^{(1)} - 
\varepsilon_{\kappa}^{(2)}} \Phi_{\vk}^{(1)}(\vr)
\label{shift-2}
\ee
where we have defined 
\be
G(\vk) = -q \kappa \sin\vartheta_{\kappa} \cos\vartheta_{\kappa} + F(\vk)
\ee
as well as
\bea
F(\vk) &=& \sum_b {\rm sign}(b)  \ID{k_1}  \; \theta(\varepsilon_F - 
\varepsilon_{\vk_1}^{(b)})\nn \\ 
&& \frac{4 \pi }{|\vk - \vk_1|^2} \sin(\vartheta_{\kappa_1} - 
\vartheta_{\kappa}) \cos(\vartheta_{\kappa_1} - \vartheta_{\kappa}) \; .
\label{func-f}
\eea
Inserting the orbital shifts (Eqs.~(\ref{shift-1}) and (\ref{shift-2})) 
as well as the orbitals (Eqs.~(\ref{orb-1}) and (\ref{orb-2})) it is 
straigthforward to see that the first OEP equation (\ref{oep-s1}) is 
satisfied by our ansatz, i.e., 
\be
\sum_b \sum_{\vk} \; \theta(\varepsilon_F - \varepsilon_{\vk}^{(b)}) 
\left( \Phi_{\vk}^{(b) \dagger}(\vr) \Psi_{\vk}^{(b)}(\vr) + 
c.c. \right) = 0 \; . 
\ee
This can easily be understood from the physical content of the OEP 
equations: if we start from the KS Hamiltonian as non-interacting reference 
and perform a perturbation expansion of the interacting density in the 
perturbation $\hat{W}_{Clb} - \hat{V}_{xc} - \mu_B \bfsigma \hat{\vB}_{xc}$, 
where $\hat{W}_{Clb}$ is the operator of the electron-electron interaction 
and $\hat{V}_{xc} + \mu_B \bfsigma \hat{\vB}_{xc}$ is the operator of the 
KS exchange-correlation potentials, then the OEP equation in EXX 
simply says that the density remains unchanged to first order. In our case 
we keep the density fixed and therefore the OEP equation (\ref{oep-s1}) holds. 

A similar argument can be used for the $z$-component of the OEP equation 
(\ref{oep-s2}) which says that the $z$-component $m(\vr)$ of the 
magnetization density remains unchanged under the same perturbation to first 
order. This equation reads explicitly
\bea
\lefteqn{
\sum_b \sum_{\vk} \; \theta(\varepsilon_F - \varepsilon_{\vk}^{(b)}) 
\left( \Phi_{\vk}^{(b) \dagger}(\vr) \sigma_z \Psi_{\vk}^{(b)}(\vr) + 
c.c. \right) } \nn\\
&=& - 2 \sum_b {\rm sign}(b)  \nn\\ &&
\ID{k} \; \theta(\varepsilon_F - \varepsilon_{\vk}^{(b)}) 
\frac{\sin\vartheta_{\kappa} \cos\vartheta_{\kappa} G(\vk)}
{\varepsilon_{\kappa}^{(1)} - \varepsilon_{\kappa}^{(2)}} = 0 
\eea
where the last equality can most easily be seen by noting that the integrand 
is an odd function under the transformation $\kappa \to -\kappa$ and all the 
integrals are over a symmetric range around $\kappa=0$. 

For the $x-$ and $y-$component of Eq.~(\ref{oep-s2}) we obtain 
\bea
\sum_b \sum_{\vk} \; \theta(\varepsilon_F - \varepsilon_{\vk}^{(b)}) 
\left( \Phi_{\vk}^{(b) \dagger}(\vr) \sigma_x \Psi_{\vk}^{(b)}(\vr) + 
c.c. \right)  \nn\\
= J(\varepsilon_F,q,B) \cos(q z) = 0 
\label{oep-x}
\eea
and 
\bea
\sum_b \sum_{\vk} \; \theta(\varepsilon_F - \varepsilon_{\vk}^{(b)}) 
\left( \Phi_{\vk}^{(b) \dagger}(\vr) \sigma_x \Psi_{\vk}^{(b)}(\vr) + 
c.c. \right) \nn\\
= J(\varepsilon_F,q,B) \sin(q z) = 0 
\label{oep-y}
\eea
where 
\bea
\lefteqn{
J(\varepsilon_F,q,B) = 2 \sum_b  {\rm sign}(b)} \nn\\
&&  \ID{k} \; \theta(\varepsilon_F - \varepsilon_{\vk}^{(b)}) 
\frac{\left(\cos^2\vartheta_{\kappa} - \sin^2\vartheta_{\kappa} 
\right) G(\vk)}{\varepsilon_{\kappa}^{(1)} - \varepsilon_{\kappa}^{(2)}}
\label{func-j}
\eea
Since Eqs.~(\ref{oep-x}) and (\ref{oep-y}) have to be satisfied for all 
values of $z$, we obtain only the condition
\be
J(\varepsilon_F,q,B) = 0 \; , 
\label{oep-xy}
\ee
i.e., the two OEP equations are not independent. 

\begin{figure}[tb]
\includegraphics[width=.45\textwidth]{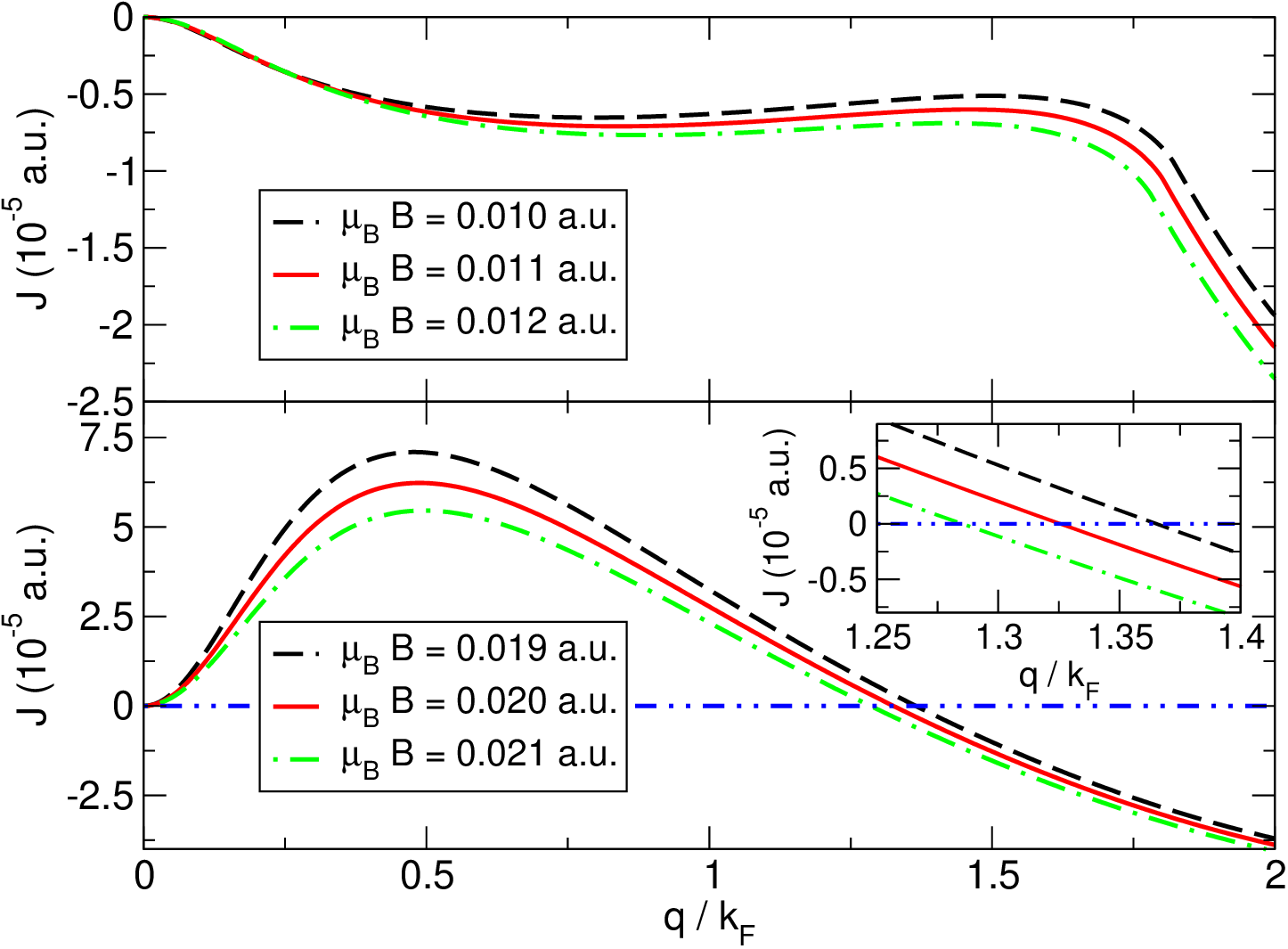}
\caption{Upper panel: $J$ of Eq.~(\protect\ref{func-j}) for $r_s=5.4$ as 
function of $q/k_F$ and different values of $B$ for the case with occupations 
in two bands. The parameter values are the same as in 
Fig.~\protect\ref{toten_rs5p4} for which a minimum in the total energy per 
particle was found. Since $J$ never crosses zero, in this case the minimum 
of the total energy is not consistent with the solution of the OEP equation. 
Lower panel: same as above but now for occupied single particle states only 
in the lower band. The parameter values are the same as in 
Fig.~\protect\ref{toten_rs5p4_one}. In contrast to the two-band case, now $J$ 
not only crosses zero but also does so at those values of $q/k_F$ for 
which a local minimum was found in Fig.~\protect\ref{toten_rs5p4_one} (see 
inset for magnification around the intersections with the zero axiss). 
Therefore, the OEP equation in this case is consistent with the minimization 
of the total energy per particle.}
\label{oep_rs5p4}
\end{figure}

In Fig.~\ref{oep_rs5p4} we show $J$ of Eq.~(\ref{func-j}) for $r_s=5.4$ as 
function of $q/k_F$ for different values of $B$ both for the case of 
occupations in both bands (upper panel) as well as for occupations restricted 
to the lower band (lower panel). Note that for the latter case the sum 
over bands $b$ both in Eq.~(\ref{func-j}) as well as in Eq.~(\ref{func-f}) 
only extends over the lower band, $b=1$. 

In the upper panel of Fig.~\ref{oep_rs5p4} we choose the same 
values for the parameter $B$ as used in Fig.~\ref{toten_rs5p4} which all had 
local minima for some value of $q < 2 k_F$. For these values of $B$, however, 
Eq.~(\ref{oep-xy}) is not satisfied for any value of $q$ in that range. We 
therefore have to conclude that in the two-band case the energy minimization 
is {\em not consistent with the OEP equations}.

In the lower panel of Fig.~\ref{oep_rs5p4} where only single-particle states 
of the lower band are occupied we choose the parameters as in 
Fig.~\ref{toten_rs5p4_one}. In this case, $J$ not only crosses zero  
but also does so exactly for those values of $q/k_F$ for which we found 
local minima in the total energy per particle in Fig.~\ref{oep_rs5p4} (see 
inset for a magnification of the region where $J$ crosses zero). We therefore 
conclude that in the one-band case the minimization of the total energy 
is consistent with the OEP equation, i.e., our ansatz is selfconsistent in 
this case. Again we emphasize that the resulting Slater determinant is 
{\em not} the ground state of the KS system. 

It has been shown \cite{BachLiebLossSolovej:94} that in unrestricted HF theory 
all the single-particle levels are fully occupied up to the Fermi energy. To 
the best of our knowledge, a similar statement has not been proven for 
DFT (even in EXX approximation) and our results indicate that it might not 
be true in EXX. On the other hand, the proof of 
Ref.~\onlinecite{BachLiebLossSolovej:94} holds for the true, unrestricted HF 
ground state while in our case we have restricted the symmetry 
of our problem to the SDW symmetry. It is quite conceivable that the fact 
that we find an excited-state Slater determinant as energy-minimizing 
wavefunction hints towards an instability of the SDW phase against further 
reduction of the symmetry.

\section{Summary and Conclusions}
\label{conclus}

We have investigated the SDW state of the uniform electron gas within the 
EXX approximation of noncollinear SDFT. While in the Hartree-Fock approximation 
the SDW state is energetically more stable than the paramagnetic state for 
all values of $r_s$, in EXX this is only true for values of $r_s$ larger than 
a critical value. Using an explicit ansatz for the spinor orbitals 
in the SDW state, we have performed the energy minimization of the EXX total 
energy in two ways: (i) in the first case we used as non-interacting reference 
wavefunction a {\em ground state} Slater determinant with occupied 
single-particle orbitals belonging to both single-particle energy bands, as 
long as their energy is below the Fermi energy. Then the SDW phase is 
more stable than both paramagnetic and ferromagnetic phases for 
$5.0 \stackrel{<}{\sim} r_s \stackrel{<}{\sim} 5.46$. (ii) In the second case 
we required all the occupied single-particle orbitals in the Slater determinant 
to belong to the lower band. The 
minimizing Slater determinant in this case turns out to be an excited 
state, since orbitals with orbital energies below the Fermi energy belonging 
to the second band remain unoccupied. Nevertheless, for a given $r_s$ the 
total energies of the minizing SDW states are significantly lower than in 
case (i). The range of stabilty of the SDW phase with respect to both 
paramagnetic and ferromagnetic phases is extended to $4.78 \stackrel{<}{\sim} 
r_s \stackrel{<}{\sim} 5.54$.

We then have investigated if the self-consistency conditions provided by the 
OEP equations for non-collinear SDFT are satisfied with our ansatz for the 
single-particle orbitals. We have found that for case (i) the parameter 
values minimizing the EXX total energy are {\em not} consistent with a solution 
of the OEP equations. In case (ii), on the other hand, for the same parameter 
values for which the EXX total energy is minimized also the OEP equations 
are satisfied. In this case, the solution we found is therefore selfconsistent. 

\acknowledgments

We would like to acknowledge useful discussions with Ilya Tokatly, Giovanni 
Vignale, and Nicole Helbig. We acknowledge funding by the "Grupos Consolidados 
UPV/EHU del Gobierno Vasco" (IT-319-07).

\end{document}